\documentclass[aps,prd,10pt,showkeys,superscriptaddress,nofootinbib,longbibliography]{revtex4-2}
\usepackage[T1]{fontenc}
\usepackage{amsmath,amssymb,bm,microtype}
\usepackage[colorlinks=true,citecolor=blue,linkcolor=blue,urlcolor=blue]{hyperref}
\hypersetup{pdftitle={Metric perturbations in Einstein-Hilbert gravity: An alternative perturbative formulation},pdfauthor={Sergey Bondarenko and Raghvendra Singh}}
\newcommand{\dd}{\mathrm d}
\newcommand{\tr}{\operatorname{tr}}

\newcommand{\Id}{\mathbb I}
\newcommand{\nab}{\nabla}
\newcommand{\Boxz}{\Box_0}
\newcommand{\hh}{\gamma}
\begin{document}
\title{Metric perturbations in Einstein-Hilbert gravity: An alternative perturbative formulation}
\author{Sergey Bondarenko}
\email{sergeyb@ariel.ac.il}
\author{Raghvendra Singh}
\email{raghvendra@ariel.ac.il}
\affiliation{Ariel University, Ariel 4070000, Israel}
\begin{abstract}
We derive the ordinary gauge-fixed quadratic operators of the four-dimensional Einstein--Hilbert action in exponential and inverse-metric-density variables on an arbitrary reference background. With a single nonlinear background-harmonic gauge transported through the exact field map, the density operator is the trace-reversed Lichnerowicz operator, while the exponential operator contains an additional algebraic Ricci-curvature term. This difference arises from the nonlinear transformation of the background tadpole and vanishes on Ricci-flat backgrounds; a cosmological-constant check gives the corresponding cancellation at Einstein saddles. The construction retains the exact determinant contribution and specifies the associated Faddeev--Popov operator. At the first interacting order on a flat background, we show that the two gauge-fixed cubic actions differ by a term proportional to the free field equation, and independently evaluate the same nonzero three-point coefficient in both variables, with its normalization matched to the Einstein-gravity interaction. An exact Schwarzschild configuration satisfies the same background-harmonic gauge and is stationary under unrestricted exterior variations of the gauge-fixed action. Its density and exponential reconstructions are checked together with the bulk and boundary terms; a separate radial variation illustrates the loss of an equation after imposing a restricted ansatz. These results provide an explicit comparison of established field coordinates at the classical and tree levels; quantum loop calculations additionally require a common measure, regulator and source prescription.
\end{abstract}

\maketitle
\section{Introduction}\label{App}
The choice of metric variables determines how the infinitely many interactions of Einstein--Hilbert gravity appear in a perturbative calculation. Nonlinear field redefinitions reorganize these interactions, but a comparison on a curved reference background must also retain the first variation of the action and specify how the gauge functional is transported. We study this comparison for the trace-reversed exponential variable $v$ and the inverse-metric-density fluctuation $\sigma_1$ in four-dimensional Einstein gravity, taking the cosmological constant to vanish except for a short quadratic consistency check.

The density representation has a long history~\cite{Faddeev}. Grishchuk, Petrov and Popova developed an exact field formulation on an arbitrary reference spacetime~\cite{Grish}. Cheung and Remmen obtained simplified action representations, including a cubic formulation with an auxiliary field and tree-level recursion relations; their construction also covers arbitrary curved backgrounds~\cite{Cheung1,Cheung2}. Cho, Kim and Lee derived general-order expansions in density and metric variables, with amplitude checks and a discussion of curved-background extensions~\cite{Korean}. Gauge, measure and parametrization dependence also enter the off-shell one-loop analysis of Ohta, Percacci and Pereira~\cite{Ohta2016}. Our focus is the explicit comparison of the curved quadratic operators and flat cubic interactions for the two specified coordinates using one transported gauge. We keep the density tadpole, exact determinant term and ghost transformation together in this comparison. The field variables, the existence of background formulations and the general field-redefinition equivalence are established ingredients. Earlier simplifications of the Einstein-gravity vertices also provide an external benchmark for the cubic normalization~\cite{BernGrant1999}.

We use a second-order formalism in which the action is expressed entirely through the metric and its derivatives. The reference metric $g_{0\mu\nu}$ need not be flat or satisfy the vacuum equations. In the density coordinate the exact action separates into the background action, a term linear in $\sigma_1$ and a nonlinear derivative contribution. This is a decomposition of one metric action.

The usual additive metric fluctuation is defined by
\begin{equation}
 g_{\mu\nu}=g_{0\mu\nu}+\hh_{\mu\nu}.
 \label{App1}
\end{equation}
The fields $v$ and $\sigma_1$ also vanish at the reference configuration, but their nonlinear relation to $\hh$ must be retained. Their exact definitions are given in Sec.~\ref{AppA}. The alternative formulation in the title refers to these coordinates for Einstein gravity.

The central quadratic comparison can be stated directly. For the background-divergence gauge at unit gauge parameter, write the degree-two action as $(2\kappa^2)^{-1}\langle u,\mathcal H_u u\rangle$, where $\kappa^2=32\pi G$ and the pairing uses $g_0$. We obtain
\begin{equation}
 \mathcal H_{\sigma_1}=P\Delta_L,\qquad
 \mathcal H_v=P\Delta_L-A.
 \label{AppIntroHessians}
\end{equation}
Here $P$ is four-dimensional trace reversal and $A$ is the algebraic Ricci operator,
\begin{equation}
 (Pu)_{\mu\nu}=u_{\mu\nu}-\tfrac12g_{0\mu\nu}g_0^{\rho\lambda}u_{\rho\lambda},
 \qquad
 (Au)_{\mu\nu}=R_{0\mu}{}^\rho u_{\rho\nu}+R_{0\nu}{}^\rho u_{\rho\mu},
 \label{AppIntroOperators}
\end{equation}
and $\Delta_L$ is the background Lichnerowicz operator defined in Eq.~\eqref{AppClich}. The $-A$ contribution is the ordinary second-derivative chain rule: the quadratic part of the field map multiplies the background tadpole. It vanishes on a Ricci-flat background. The operators in Eq.~\eqref{AppIntroHessians} are the quadratic forms of the same gauge-fixed action in the two coordinates; no background field equation has been imposed in deriving them. Section~\ref{AppDlambda} includes the cosmological volume term and shows how the same tadpole argument gives equality at an Einstein saddle.

The comparison extends to the first interacting order on a flat background. Section~\ref{AppD} displays the identity relating the cubic actions through the free field equation, including the nonlinear pullback of the gauge term. Appendix~\ref{AppenB} expands both cubic interactions directly, evaluates their nonzero three-point coefficients and matches the result to the published Einstein-gravity interaction with common external-state conventions. The Schwarzschild construction tests exact metric reconstruction beyond a weak-field expansion~\cite{Duff}. Its harmonic representative satisfies the same nonlinear gauge, and the full exterior Ricci tensor, density action and boundary current are evaluated explicitly. The two checks address the perturbative interaction and a finite classical stationary configuration of the gauge-fixed action, respectively.

The resulting formulation supplies compatible curved-background quadratic operators, an explicit ghost sector and flat cubic interactions in both coordinates. It also identifies the terms that would be lost by dropping the tadpole or changing the nonlinear gauge independently in the two schemes. The calculation is classical and tree-level. At the functional-integral level, the coordinate Jacobian, measure and external sources must be transported consistently; a regulated loop calculation requires additional choices not made here.

The paper is organized as follows. Section~\ref{AppA} gives the exact action and field definitions, and Sec.~\ref{AppB} identifies the nonlinear background-harmonic gauge and fixes the functional-integral conventions. Sections~\ref{AppC} and \ref{AppD} derive the two quadratic operators and the cubic comparison. Section~\ref{AppE} presents the harmonic Schwarzschild configuration and the independent radial variation, followed by the conclusions in Sec.~\ref{AppF}. Appendices~\ref{AppenA}--\ref{AppenC} contain the ghost derivation, detailed cubic calculation and radial curvature and variation identities, respectively.

\section{Einstein--Hilbert Lagrangian in curved space-time in terms of the background metric and perturbation field}\label{AppA}
We begin with the gravitational action with vanishing cosmological constant,
\begin{equation}
 S=-\frac1{4\chi}\int\dd^4x\sqrt{-g}\,g^{\mu\nu}R_{\mu\nu},
 \qquad \chi=4\pi G,\qquad \kappa^2=8\chi=32\pi G.
 \label{AppA1}
\end{equation}
We use units $c=\hbar=1$, signature $(+---)$, and
\begin{equation}
 R_{\mu\nu}=\partial_\rho\Gamma^\rho_{\mu\nu}
 -\partial_\nu\Gamma^\rho_{\mu\rho}
 +\Gamma^\rho_{\mu\nu}\Gamma^\lambda_{\rho\lambda}
 -\Gamma^\lambda_{\mu\rho}\Gamma^\rho_{\nu\lambda}.
 \label{AppA2}
\end{equation}
All spacetime indices are Greek. Background raising and lowering and the derivative $\nab\equiv\nabla^{(0)}$ refer to $g_0$, which is held fixed unless explicitly stated otherwise. Integrations by parts in the bulk assume compactly supported variations or boundary conditions with the appropriate gravitational boundary action.

The exponential metric variable $h_{\mu\nu}$ is defined by
\begin{equation}
 g_{\mu\nu}=g_{0\mu\rho}(e^{\mathsf h})^\rho{}_{\nu},
 \qquad \mathsf h^\mu{}_{\nu}=g_0^{\mu\rho}h_{\rho\nu},
 \qquad \frac{\sqrt{-g}}{\sqrt{-g_0}}=e^{\frac12\tr\mathsf h}.
 \label{AppA8}
\end{equation}
Thus $h$ agrees with $\hh$ only to first order. Introduce its trace reversal and the normalized inverse metric density,
\begin{align}
 v_{\mu\nu}&=h_{\mu\nu}-\tfrac12g_{0\mu\nu}\tr\mathsf h,
 &V^\mu{}_{\nu}&=v^{\mu\rho}g_{0\rho\nu},\nonumber\\
 \omega&=\frac{\sqrt{-g}}{\sqrt{-g_0}},
 &\sigma^{\mu\nu}&=\omega g^{\mu\nu}=g_0^{\mu\nu}+\sigma_1^{\mu\nu}
 =(e^{-V})^\mu{}_{\rho}g_0^{\rho\nu}.
 \label{AppA10}
\end{align}
The volume ratio makes $\sigma^{\mu\nu}$ a tensor under simultaneous coordinate transformations of the full and background metrics. The inverse matrix is denoted by $\tau_{\mu\nu}$:
\begin{equation}
 \tau_{\mu\rho}\sigma^{\rho\nu}=\delta_\mu{}^\nu,
 \qquad \tau_{\mu\nu}=g_{0\mu\rho}(e^V)^\rho{}_{\nu},
 \qquad \tau_{1\mu\nu}=\tau_{\mu\nu}-g_{0\mu\nu}.
 \label{AppA11}
\end{equation}
The inverse-density shift $\tau_1$ is not the background lowering $\sigma_{1\mu\nu}=g_{0\mu\rho}g_{0\nu\lambda}\sigma_1^{\rho\lambda}$. This distinction is necessary at every nonlinear order.

Set
\begin{equation}
 Q^\mu{}_{\nu}=\sigma^{\mu\rho}g_{0\rho\nu}=\Id+X,
 \qquad X^\mu{}_{\nu}=\sigma_1^{\mu\rho}g_{0\rho\nu},
 \qquad H=\ln\det Q.
 \label{AppA12}
\end{equation}
In four dimensions,
\begin{equation}
 \det Q=\omega^2,
 \qquad H=-\tr V=\tr\mathsf h,
 \qquad g_{\mu\nu}=e^{H/2}\tau_{\mu\nu}.
 \label{AppA121}
\end{equation}
We use the positive volume-ratio branch with nonsingular $\sigma$ of Lorentzian inertia. The exponential and logarithm are mutually inverse in a sufficiently small neighborhood of the background; no global exponential chart for all Lorentzian metrics is assumed. The exact density reconstruction can be used for finite configurations on its nonsingular domain.

The expansions relating the two schemes are
\begin{align}
 \sigma_1^{\mu\nu}&=-v^{\mu\nu}
 +\tfrac12v^{\mu\rho}g_{0\rho\lambda}v^{\lambda\nu}+O(v^3),\nonumber\\
 \tau_{1\mu\nu}&=-\sigma_{1\mu\nu}
 +\sigma_{1\mu\rho}\sigma_1{}^\rho{}_{\nu}+O(\sigma_1^3).
 \label{AppA123}
\end{align}
Both are field redefinitions of the same metric. The inverse in Eq.~\eqref{AppA11}, rather than a truncated series, defines a finite density configuration.

To write the action covariantly, introduce the difference of connections $C^\rho{}_{\mu\nu}=\Gamma^\rho{}_{\mu\nu}[g]-\Gamma^\rho{}_{\mu\nu}[g_0]$. Direct substitution of Eq.~\eqref{AppA121} gives
\begin{align}
 C^\rho{}_{\mu\nu}={}&\tfrac12\sigma^{\rho\lambda}
 (\nab_\mu\tau_{\lambda\nu}+\nab_\nu\tau_{\lambda\mu}-\nab_\lambda\tau_{\mu\nu})\nonumber\\
 &+\tfrac14(\delta^\rho_\mu H_\nu+\delta^\rho_\nu H_\mu
 -\sigma^{\rho\lambda}\tau_{\mu\nu}H_\lambda),\qquad H_\mu=\partial_\mu H.
 \label{AppAconnection}
\end{align}
The steps in reducing the action are useful for fixing the sign of the determinant term. The curvature-difference identity is
\begin{align}
 R_{\mu\nu}[g]={}&R_{0\mu\nu}
 +\nab_\rho C^\rho{}_{\mu\nu}-\nab_\nu C^\rho{}_{\mu\rho}
 +C^\rho{}_{\rho\lambda}C^\lambda{}_{\mu\nu}
 -C^\rho{}_{\nu\lambda}C^\lambda{}_{\mu\rho}.
 \label{AppAcurvaturedifference}
\end{align}
Metric compatibility of the full metric and $\partial_\rho\ln\omega=C^\lambda{}_{\rho\lambda}$ imply
\begin{align}
 \nab_\rho\sigma^{\mu\nu}
 &=-C^\mu{}_{\rho\lambda}\sigma^{\lambda\nu}
   -C^\nu{}_{\rho\lambda}\sigma^{\mu\lambda}
   +C^\lambda{}_{\rho\lambda}\sigma^{\mu\nu},
 & C^\lambda{}_{\rho\lambda}&=\tfrac12 H_\rho.
 \label{AppAcompatibility}
\end{align}
Multiplying Eq.~\eqref{AppAcurvaturedifference} by $\sigma^{\mu\nu}$ and integrating its two derivative terms by parts gives a divergence and a quadratic connection expression. In particular, the derivatives acting on $\sigma$ are evaluated with Eq.~\eqref{AppAcompatibility}; they must not be discarded. The result is
\begin{equation}
 \omega R[g]=\mathcal L_\sigma+\nab_\rho W^\rho,
 \qquad W^\rho=\sigma^{\mu\nu}C^\rho{}_{\mu\nu}-\sigma^{\mu\rho}C^\nu{}_{\mu\nu},
 \label{AppAboundary}
\end{equation}
where the bulk Lagrangian is~\cite{Cheung1,Korean}
\begin{align}
 \mathcal L_\sigma={}&\sigma^{\mu\nu}R_{0\mu\nu}
 +(\nab_\mu\sigma^{\alpha\beta})(\nab_\nu\tau_{\rho\beta})
 \left(\tfrac14\sigma^{\mu\nu}\delta_\alpha^\rho
       -\tfrac12\sigma^{\mu\rho}\delta_\alpha^\nu\right)
 +\tfrac18\sigma^{\mu\nu}H_\mu H_\nu.
 \label{AppA1401}
\end{align}
For clarity, the connection expression and its reduction to density derivatives are
\begin{align}
 \mathcal L_{\mathrm{der}}
 &\equiv\sigma^{\mu\nu}
 (C^\rho{}_{\mu\lambda}C^\lambda{}_{\nu\rho}
  -C^\rho{}_{\mu\nu}C^\lambda{}_{\rho\lambda})\nonumber\\
 &=-\tfrac14\sigma^{\mu\nu}\tau_{\alpha\rho}\tau_{\beta\lambda}
       (\nab_\mu\sigma^{\alpha\beta})(\nab_\nu\sigma^{\rho\lambda})
   +\tfrac12\tau_{\alpha\rho}
       (\nab_\lambda\sigma^{\alpha\beta})(\nab_\beta\sigma^{\lambda\rho})
   +\tfrac18\sigma^{\mu\nu}H_\mu H_\nu.
 \label{AppAdensitycontractions}
\end{align}
This follows by inserting Eq.~\eqref{AppAconnection} and using
\begin{equation}
 \nab_\mu\tau_{\alpha\beta}
 =-\tau_{\alpha\rho}\tau_{\beta\lambda}\nab_\mu\sigma^{\rho\lambda},
 \qquad H_\mu=\tau_{\alpha\beta}\nab_\mu\sigma^{\alpha\beta}.
 \label{AppAinverseidentities}
\end{equation}
In the second term of Eq.~\eqref{AppAdensitycontractions}, one inverse-density factor has canceled against $\sigma$. Reversing that cancellation and using the first identity in Eq.~\eqref{AppAinverseidentities} gives precisely the two derivative contractions in Eq.~\eqref{AppA1401}. This derivation fixes the bulk action together with its boundary current.

In particular, $\frac12\sigma^{\mu\nu}\partial_\mu\ln\omega\,\partial_\nu\ln\omega=\frac18\sigma^{\mu\nu}H_\mu H_\nu$. The coefficient is $1/8$ when $H=\tr\mathsf h$, not $1/2$. A boundary term cannot independently reverse this sign. For $g=e^{2\psi}g_0$, Eq.~\eqref{AppA1401} reduces to $e^{2\psi}[R_0+6(\nab\psi)^2]$, consistently with the Einstein--Hilbert conformal transformation after integration by parts.

The background split now takes the simple form
\begin{equation}
 S_{\mathrm{bulk}}=S[g_0]
 -\frac1{4\chi}\int\dd^4x\sqrt{-g_0}\,
 \left[\sigma_1^{\mu\nu}R_{0\mu\nu}+\mathcal L_{\mathrm{der}}[\sigma_1;g_0]\right].
 \label{AppA15}
\end{equation}
Here $\mathcal L_{\mathrm{der}}$ is the derivative part of Eq.~\eqref{AppA1401}. It begins quadratically in $\sigma_1$ and contains infinitely many interactions through $\tau$ and $H$. The linear term is the background tadpole. Covariant derivative commutators still supply curvature terms in the quadratic action. Equation~\eqref{AppA15} separates contributions of a single metric action; neither the nonlinear derivative terms nor a coordinate-dependent part of them constitute a separate free classical theory.

\section{Single-variable action, gauge fixing and functional determinants}\label{AppB}
The trace-dependent term in Eq.~\eqref{AppA1401} can already be written entirely in terms of the density fluctuation:
\begin{equation}
 H=\ln\det(\Id+\sigma_1 g_0),
 \qquad \partial_\mu H=\tr[(\Id+X)^{-1}\nab_\mu X].
 \label{AppBtrace}
\end{equation}
Thus the action is a single-variable functional before any gauge fixing. Alternatively, inserting $\sigma_1(v)$ gives its exponential representation. Since $v$ and $\sigma_1$ are related by Eq.~\eqref{AppA10}, they cannot be integrated over as independent variables to cancel the trace term. An auxiliary identity does not justify deleting a field-dependent contribution from the action.

For perturbation theory choose one four-component gauge condition,
\begin{equation}
 F^\mu[\sigma_1]=\nab_\nu\sigma_1^{\mu\nu},
 \qquad S_{\mathrm{GF}}=\frac1{8\chi\alpha}\int\dd^4x\sqrt{-g_0}\,
 g_{0\mu\nu}F^\mu F^\nu.
 \label{AppBgauge}
\end{equation}
The linear metric relation is $\hh_{\mu\nu}=-\sigma_{1\mu\nu}+\frac12g_{0\mu\nu}\sigma_1+O(\sigma_1^2)$, where $\sigma_1=g_{0\mu\nu}\sigma_1^{\mu\nu}$. Hence $F$ is the de Donder condition at linear order, up to an overall sign. We use $\alpha=1$ below. No trace constraint or second gauge condition is imposed. In the first perturbative scheme the same gauge functional means $F[\sigma_1(v)]$, including its nonlinear terms.

The geometric meaning of this condition is exact. With the connection difference $C$ of Eq.~\eqref{AppAconnection}, metric compatibility and $\partial_\nu\ln\omega=C^\rho{}_{\rho\nu}$ imply
\begin{equation}
 F^\mu=\nab_\nu\sigma^{\mu\nu}
 =-\omega g^{\rho\lambda}C^\mu{}_{\rho\lambda}
 =-\omega g^{\rho\lambda}
   \bigl(\Gamma^\mu_{\rho\lambda}[g]-\Gamma^\mu_{\rho\lambda}[g_0]\bigr).
 \label{AppBharmonic}
\end{equation}
Indeed, contracting Eq.~\eqref{AppAcompatibility} cancels its connection-trace terms. Since $\nab g_0^{-1}=0$, the divergences of $\sigma$ and $\sigma_1$ coincide. Thus $F=0$ is the nonlinear background-harmonic (wave-map) condition relative to $g_0$. For a flat Cartesian reference metric it becomes
\begin{equation}
 \partial_\nu(\sqrt{-g}\,g^{\mu\nu})=0
 \quad\Longleftrightarrow\quad
 g^{\rho\lambda}\Gamma^\mu_{\rho\lambda}[g]=0
 \quad\Longleftrightarrow\quad \Box_g x^\mu=0.
 \label{AppBharmonicCartesian}
\end{equation}
This is the established harmonic-coordinate condition for the gothic inverse metric~\cite{Blanchet2014}. Equation~\eqref{AppBharmonic} is covariant under simultaneous coordinate transformations of $g$ and $g_0$; the middle expression in Eq.~\eqref{AppBharmonicCartesian} alone is a condition on a particular coordinate chart. The reference connection must be retained in general coordinates. We keep precisely the gauge-fixing weight in Eq.~\eqref{AppBgauge}: replacing $F$ by $-F/\omega$ preserves its zero set but changes the nonlinear gauge-fixing action and the off-shell ghost operator.

The Faddeev--Popov operator is defined by varying this gauge under a quantum diffeomorphism at fixed background~\cite{Faddeev,FaddeevPopov1967}:
\begin{equation}
 (M[\sigma_1]\xi)^\mu=\delta_\xi F^\mu[\sigma_1],
 \qquad M[0]^\mu{}_{\rho}=-\Boxz\delta^\mu{}_{\rho}-R_0{}^\mu{}_{\rho}.
 \label{AppBFP}
\end{equation}
Here $\Boxz=g_0^{\mu\nu}\nab_\mu\nab_\nu$. Appendix~\ref{AppenA} gives the full operator, its ghost action and the first interaction explicitly. Inverses require boundary or Feynman prescriptions. Residual transformations and background isometries, if present, must be treated separately; no globally unique gauge representative is assumed.

\subsection{Local field redefinitions, Jacobian factors, and ghost determinants}\label{AppBsubA}
For a chosen starting measure $\mu[\sigma_1]\mathcal D\sigma_1$, the formal coordinate change gives
\begin{align}
 Z[J]=\int\mathcal Dv\mathcal D\bar c\mathcal Dc\,
 \mu[\sigma_1(v)]\mathcal J[v]\,
 \exp\bigl\{i\bigl(&S[\sigma_1(v)]+S_{\mathrm{GF}}[\sigma_1(v)]
 +S_{\mathrm{gh}}[\sigma_1(v),\bar c,c]
 +J\cdot\sigma_1(v)\bigr)\bigr\},\nonumber\\
 \mathcal J[v]&=\left|\operatorname{Det}_{\mathrm{Sym}}
 \frac{\delta\sigma_1}{\delta v}\right|.
 \label{AppBmeasure}
\end{align}
The equality presumes the same regulated integration problem, with its source and any defining restrictions transported. The map is local and has derivative $-\Id$ on symmetric components at the background. Its functional Jacobian nevertheless needs a measure and regularization prescription. It is not the Faddeev--Popov determinant, which acts on vector gauge parameters. Replacing $J\cdot\sigma_1(v)$ by a source linear in $v$ changes the off-shell functional. These distinctions matter for loop calculations and off-shell effective actions~\cite{Ohta2016}; they do not alter the classical field redefinition or the tree-level comparison made here. We do not evaluate a regulated functional determinant in this paper.

\section{First perturbative scheme: perturbative action in terms of the \texorpdfstring{$v^{\mu\nu}$}{v} field}\label{AppC}
In the first scheme the independent symmetric field is $v^{\mu\nu}$. The action is obtained by substituting $\sigma=e^{-V}g_0^{-1}$, $\tau=g_0e^V$ and $H=-\tr V$ into Eq.~\eqref{AppA1401}. The gauge and ghost actions are pulled back through the same map. The series needed through cubic order is
\begin{align}
 \sigma_1^{\mu\nu}&=-v^{\mu\nu}
  +\tfrac12(vg_0v)^{\mu\nu}
  -\tfrac16(vg_0vg_0v)^{\mu\nu}+O(v^4),\nonumber\\
 \tau_{1\mu\nu}&=v_{\mu\nu}
  +\tfrac12(vg_0^{-1}v)_{\mu\nu}
  +\tfrac16(vg_0^{-1}vg_0^{-1}v)_{\mu\nu}+O(v^4).
 \label{AppC3}
\end{align}
Matrix products in the first line use contravariant $v$, and those in the second use its background lowering. In particular the trace term in the action remains $\sigma^{\mu\nu}\partial_\mu\tr V\,\partial_\nu\tr V/8$.

\subsection{Linear and quadratic terms on a curved reference metric}
Write $S^{(n)}$ for the term homogeneous of degree $n$, including its Taylor factor. The term linear in $v$ is
\begin{equation}
 S_v^{(1)}=\frac1{4\chi}\int\dd^4x\sqrt{-g_0}\,
       v^{\mu\nu}R_{0\mu\nu}.
 \label{AppC4}
\end{equation}
It vanishes for a vacuum reference metric. An expansion about a general background retains this term; such a background is not an unsourced stationary point.

For a symmetric tensor $u_{\mu\nu}$ define
\begin{equation}
 (Pu)_{\mu\nu}=u_{\mu\nu}-\tfrac12g_{0\mu\nu}u,
 \quad u=g_0^{\rho\lambda}u_{\rho\lambda},\qquad
 \langle u,w\rangle=\int\dd^4x\sqrt{-g_0}\,u^{\mu\nu}w_{\mu\nu}.
 \label{AppCtrace}
\end{equation}
In four dimensions $P^2=\Id$. With the curvature convention of Eq.~\eqref{AppA2}, equivalently $[\nab_\mu,\nab_\nu]a^\rho=R_0{}^\rho{}_{\lambda\mu\nu}a^\lambda$, the Lichnerowicz operator is
\begin{equation}
 (\Delta_Lu)_{\mu\nu}=-\Boxz u_{\mu\nu}
 -2R_{0\mu\rho\nu\sigma}u^{\rho\sigma}
 +R_{0\mu}{}^\rho u_{\rho\nu}+R_{0\nu}{}^\rho u_{\rho\mu}.
 \label{AppClich}
\end{equation}
Trace reversal commutes with this operator. At $\alpha=1$ the gauge-fixed quadratic action is
\begin{align}
 (S+S_{\mathrm{GF}})_v^{(2)}
 &=\frac1{16\chi}\langle v,P\Delta_Lv\rangle
 -\frac1{8\chi}\int\dd^4x\sqrt{-g_0}\,
 R_{0\mu\nu}v^{\mu\rho}g_{0\rho\lambda}v^{\lambda\nu}\nonumber\\
 &=\frac1{16\chi}\langle v,\mathcal H_vv\rangle,
 \qquad \mathcal H_v=P\Delta_L-A,\nonumber\\
 (Au)_{\mu\nu}&=R_{0\mu}{}^\rho u_{\rho\nu}
                     +R_{0\nu}{}^\rho u_{\rho\mu}.
 \label{AppC7}
\end{align}
The first term follows from the quadratic derivative action, derived in Sec.~\ref{AppD}; the second follows from the quadratic term in $\sigma_1(v)$ multiplying $R_{0\mu\nu}$ in Eq.~\eqref{AppA15}. It would be lost by dropping the density tadpole before changing variables. Equivalently, after integration by parts,
\begin{equation}
 (S+S_{\mathrm{GF}})_v^{(2)}
 =\frac1{16\chi}\int\dd^4x\sqrt{-g_0}
 \left[(\nab_\lambda v_{\mu\nu})(\nab^\lambda v^{\mu\nu})
       -\tfrac12(\nab_\lambda v)(\nab^\lambda v)
       -2R_{0\mu\rho\nu\sigma}v^{\mu\nu}v^{\rho\sigma}\right].
 \label{AppC8}
\end{equation}
Here $v=g_{0\mu\nu}v^{\mu\nu}$. The Ricci contributions in $\Delta_L$ cancel the explicit term inherited from the tadpole, but the Riemann-curvature coupling remains. Equations~\eqref{AppC7}--\eqref{AppC8} are background-covariant quadratic forms, rather than a flat-space replacement of a curved kinetic term.

\subsection{Propagator and interactions}
The inverse of the quadratic form, with an appropriate boundary prescription, determines the propagator. In a Cartesian flat reference metric, $\mathcal H_v=P(-\Box)$ and the Feynman propagator for the unnormalized field is
\begin{align}
 \langle v^{\mu\nu}(p)v^{\rho\sigma}(q)\rangle_0
 &=(2\pi)^4\delta^{(4)}(p+q)
 \frac{8\chi\,i}{p^2+i0}P^{\mu\nu,\rho\sigma},\nonumber\\
 P^{\mu\nu,\rho\sigma}
 &=\tfrac12(\eta^{\mu\rho}\eta^{\nu\sigma}
           +\eta^{\mu\sigma}\eta^{\nu\rho}
           -\eta^{\mu\nu}\eta^{\rho\sigma}).
 \label{AppC13}
\end{align}
The Fourier convention is $v(x)=\int\dd^4p\,(2\pi)^{-4}v(p)e^{ip\cdot x}$. The trace term in $P$ is necessary to invert the Einstein--Hilbert quadratic form. Rescaling $v$ by $\kappa=\sqrt{8\chi}$ gives the usual canonical normalization. On a curved background the inverse is instead that of $\mathcal H_v$, with its curvature terms and prescribed boundary conditions; no translation-invariant momentum formula is implied.

The interaction terms follow by expanding the complete action, including the trace term and the pullback $F[\sigma_1(v)]$ of the gauge functional. The latter is nonlinear in $v$ even though $F[\sigma_1]$ is linear in $\sigma_1$. The same substitution generates the ghost interactions from Appendix~\ref{AppenA}. Appendix~\ref{AppenB} displays the cubic comparison and an explicit three-point check. This construction reorganizes the interactions of Einstein gravity without introducing an additional field.

\section{Second perturbative scheme: perturbative action in terms of the \texorpdfstring{$\sigma_1^{\mu\nu}$}{sigma1} field}\label{AppD}
In the second scheme $\sigma_1^{\mu\nu}$ is independent, and the full action is Eq.~\eqref{AppA1401} with
\begin{equation}
 \tau=(g_0^{-1}+\sigma_1)^{-1},\qquad
 H=\ln\det(\Id+X),\qquad X=\sigma_1g_0.
 \label{AppD1}
\end{equation}
The perturbative series is
\begin{align}
 \tau_{1\mu\nu}&=-\sigma_{1\mu\nu}
  +\sigma_{1\mu\rho}\sigma_1{}^\rho{}_{\nu}
  -\sigma_{1\mu\rho}\sigma_1{}^\rho{}_{\lambda}\sigma_1{}^\lambda{}_{\nu}
  +O(\sigma_1^4),\nonumber\\
 H&=\tr X-\tfrac12\tr X^2+\tfrac13\tr X^3+O(X^4).
 \label{AppD6}
\end{align}
The inverse and logarithm in Eq.~\eqref{AppD1} define the exact functional. A truncation of Eq.~\eqref{AppD6} defines only a chosen perturbative order.

\subsection{Quadratic action and its relation to the first scheme}
The first variation can be obtained directly from the inverse-metric variation. At fixed background,
\begin{equation}
 \delta\sigma^{\mu\nu}=\omega\left(\delta g^{\mu\nu}
 -\tfrac12g^{\mu\nu}g_{\rho\lambda}\delta g^{\rho\lambda}\right),
 \qquad
 \delta S=-\frac1{4\chi}\int\dd^4x\sqrt{-g_0}\,
 R_{\mu\nu}[g]\delta\sigma^{\mu\nu}.
 \label{AppDvariation}
\end{equation}
To see the cancellation of the scalar-curvature term explicitly, take the trace with $\tau_{\mu\nu}$ and invert the first identity:
\begin{equation}
 \delta g^{\mu\nu}=\omega^{-1}
 \left(\delta\sigma^{\mu\nu}
       -\tfrac12\sigma^{\mu\nu}\tau_{\rho\lambda}\delta\sigma^{\rho\lambda}\right).
 \label{AppDinversevariation}
\end{equation}
Since $G^\rho{}_{\rho}=-R$ in four dimensions, the Einstein--Hilbert variation becomes
\begin{equation}
 \sqrt{-g}\,G_{\mu\nu}\delta g^{\mu\nu}
 =\sqrt{-g_0}\left(G_{\mu\nu}-\tfrac12g_{\mu\nu}G^\rho{}_{\rho}\right)
       \delta\sigma^{\mu\nu}
 =\sqrt{-g_0}\,R_{\mu\nu}\delta\sigma^{\mu\nu}.
 \label{AppDtracecancellation}
\end{equation}
This proves the second identity in Eq.~\eqref{AppDvariation}. At the reference configuration,
\begin{equation}
 S_{\sigma_1}^{(1)}=-\frac1{4\chi}\int\dd^4x\sqrt{-g_0}\,
 \sigma_1^{\mu\nu}R_{0\mu\nu}.
 \label{AppDtadpole}
\end{equation}

For completeness the quadratic operator follows from the ordered Palatini identity for a metric variation $p_{\mu\nu}$:
\begin{equation}
 \delta R_{\mu\nu}[p]=\tfrac12\left(
 \nab_\rho\nab_\mu p^\rho{}_{\nu}
 +\nab_\rho\nab_\nu p^\rho{}_{\mu}
 -\Boxz p_{\mu\nu}-\nab_\mu\nab_\nu p\right).
 \label{AppDpalatini}
\end{equation}
Here $p=g_0^{\mu\nu}p_{\mu\nu}$. The required commutator on a mixed tensor is
\begin{equation}
 [\nab_\rho,\nab_\mu]p^\rho{}_{\nu}
 =R_{0\lambda\mu}p^\lambda{}_{\nu}
  -R_0{}^\lambda{}_{\nu\rho\mu}p^\rho{}_{\lambda}.
 \label{AppDpalatinicommutator}
\end{equation}
Using it in the two ordered terms in Eq.~\eqref{AppDpalatini} yields
\begin{equation}
 \delta R_{\mu\nu}[p]
 =\tfrac12\left[(\Delta_Lp)_{\mu\nu}
 +\nab_\mu\nab^\rho p_{\rho\nu}
 +\nab_\nu\nab^\rho p_{\rho\mu}
 -\nab_\mu\nab_\nu p\right].
 \label{AppDcommutedpalatini}
\end{equation}
The linear density-to-metric relation $p=-P\sigma_1$ gives
$\nab^\rho p_{\rho\nu}=-F_\nu+\frac12\nab_\nu\sigma_1$ and $p=\sigma_1$.
The scalar double derivatives therefore cancel. Moreover,
$\Delta_L(g_0 f)=-g_0\Boxz f$ and
$\operatorname{tr}(\Delta_Lu)=-\Boxz\operatorname{tr}u$, so that
$[P,\Delta_L]=0$ on any reference background. Hence
\begin{equation}
 \delta R_{\mu\nu}[-P\sigma_1]
 =-\tfrac12\left[(P\Delta_L\sigma_1)_{\mu\nu}
                  +\nab_\mu F_\nu+\nab_\nu F_\mu\right].
 \label{AppDricci}
\end{equation}
Consequently the quadratic Einstein--Hilbert action and its gauge-fixed form are
\begin{align}
 S_{\sigma_1}^{(2)}&=\frac1{16\chi}\langle\sigma_1,P\Delta_L\sigma_1\rangle
                   -\frac1{8\chi}\langle F,F\rangle,\nonumber\\
 (S+S_{\mathrm{GF}})_{\sigma_1}^{(2)}
 &=\frac1{16\chi}\langle\sigma_1,\mathcal H_{\sigma_1}\sigma_1\rangle,
 \qquad \mathcal H_{\sigma_1}=P\Delta_L.
 \label{AppD6c}
\end{align}
The pairing for $F$ is the corresponding vector pairing with $g_0$. As a check, direct expansion of the bulk Lagrangian gives
\begin{equation}
 \mathcal L_\sigma^{(2)}
 =-\tfrac14(\nab_\mu\sigma_1^{\alpha\beta})(\nab^\mu\sigma_{1\alpha\beta})
 +\tfrac12(\nab_\mu\sigma_1^{\alpha\beta})(\nab_\alpha\sigma_1{}^\mu{}_{\beta})
 +\tfrac18(\nab_\mu\sigma_1)(\nab^\mu\sigma_1).
 \label{AppD6a}
\end{equation}
After integration by parts, the difference between the mixed derivative term and $F^2$ is
$-\int\dd^4x\sqrt{-g_0}\,\sigma_1^{\alpha\beta}
[\nab_\mu,\nab_\alpha]\sigma_1{}^\mu{}_{\beta}$.
Evaluating the commutator as in Eq.~\eqref{AppDpalatinicommutator} gives
\begin{align}
 \int\dd^4x\sqrt{-g_0}\left[
 (\nab_\mu\sigma_1^{\alpha\beta})(\nab_\alpha\sigma_1{}^\mu{}_{\beta})
 -F_\mu F^\mu\right]
 =\int\dd^4x\sqrt{-g_0}\left[
 R_{0\mu\rho\nu\sigma}\sigma_1^{\mu\nu}\sigma_1^{\rho\sigma}
 -R_{0\mu\nu}\sigma_1^{\mu\rho}\sigma_1{}^\nu{}_{\rho}\right].
 \label{AppDcommutator}
\end{align}
This identity supplies the curvature potential in Eq.~\eqref{AppD6c} and explains why setting $g_0=\eta$ too early misses part of the comparison.

Equations~\eqref{AppC7} and \eqref{AppD6c} imply
\begin{equation}
 \mathcal H_v=\mathcal H_{\sigma_1}-A.
 \label{AppD8a}
\end{equation}
The term $-A$ is the ordinary second-derivative chain rule applied to a nonlinear change of field coordinates: the second derivative of the map multiplies the first derivative of the action. It vanishes on a Ricci-flat reference metric. The two quadratic forms then coincide after the linear identification $\sigma_1=-v$. For a general background they are different off-shell Hessians of the same action. The source and measure qualifications in Sec.~\ref{AppB} apply to a comparison of their generating functionals.

In flat space Eq.~\eqref{AppD6c} gives the propagator \eqref{AppC13}, with $v$ replaced by $\sigma_1$. Equality of the flat propagators is therefore exact, including the trace numerator and normalization. It is not a statement that arbitrary unsourced curved-background Hessians agree.

\subsection{A cosmological-constant consistency check}
\label{AppDlambda}
The role of the tadpole can also be checked for
\begin{equation}
 S_\Lambda=-\frac1{4\chi}\int\dd^4x\sqrt{-g}\,(R-2\Lambda).
 \label{AppDlambdaAction}
\end{equation}
Only this subsection includes $\Lambda$; the flat cubic and Schwarzschild calculations below have $\Lambda=0$. In four dimensions, with $X=\sigma_1g_0$ and $V=vg_0$, the two volume expansions are
\begin{align}
 \omega(\sigma_1)&=\sqrt{\det(\Id+X)}
 =1+\tfrac12\tr X+\tfrac18(\tr X)^2-\tfrac14\tr X^2+O(X^3),\nonumber\\
 \omega(v)&=e^{-\tr V/2}
 =1-\tfrac12\tr V+\tfrac18(\tr V)^2+O(V^3).
 \label{AppDlambdaVolumes}
\end{align}
Writing $\dd\bar V=\dd^4x\sqrt{-g_0}$, the cosmological term therefore contributes
\begin{align}
 S^{(2)}_{\mathrm{cos},\sigma_1}
 &=\frac{\Lambda}{16\chi}\int\dd\bar V\,
       [(\tr X)^2-2\tr X^2],\nonumber\\
 S^{(2)}_{\mathrm{cos},v}
 &=\frac{\Lambda}{16\chi}\int\dd\bar V\,(\tr V)^2.
 \label{AppDlambdaQuadratic}
\end{align}
Define $(Tu)_{\mu\nu}=g_{0\mu\nu}g_0^{\rho\lambda}u_{\rho\lambda}$, so $P=\Id-T/2$. With the same gauge and quadratic normalization as above,
\begin{equation}
 \mathcal H_{\sigma_1,\Lambda}=P(\Delta_L-2\Lambda\Id),
 \qquad
 \mathcal H_{v,\Lambda}=P\Delta_L-A+\Lambda T.
 \label{AppDlambdaHessians}
\end{equation}
The first variation in the density coordinate is now
\begin{equation}
 S^{(1)}_{\sigma_1,\Lambda}
 =-\frac1{4\chi}\int\dd\bar V\,
   (R_{0\mu\nu}-\Lambda g_{0\mu\nu})\sigma_1^{\mu\nu}.
 \label{AppDlambdaTadpole}
\end{equation}
Its pullback through Eq.~\eqref{AppA123} supplies the quadratic contribution
\begin{equation}
 -\frac1{8\chi}\int\dd\bar V\,
 (R_{0\mu\nu}-\Lambda g_{0\mu\nu})(vg_0v)^{\mu\nu}
 =\frac1{16\chi}\langle v,(-A+2\Lambda\Id)v\rangle.
 \label{AppDlambdaPullback}
\end{equation}
This agrees with the difference of the independently expanded operators,
\begin{equation}
 \mathcal H_{v,\Lambda}-\mathcal H_{\sigma_1,\Lambda}
 =-A+2\Lambda\Id.
 \label{AppDlambdaDifference}
\end{equation}
At an Einstein saddle $R_{0\mu\nu}=\Lambda g_{0\mu\nu}$, one has $A=2\Lambda\Id$, and both operators reduce to $P(\Delta_L-2\Lambda\Id)$. The cancellation is the expected second-variation chain rule at a stationary point, and extends the Ricci-flat consistency check without introducing a new equivalence principle.

\subsection{Interactions and the flat cubic comparison}
The density gauge term is exactly quadratic, whereas the inverse density and $H$ generate higher interactions in the gravitational action. The first interacting order makes the relation between the two schemes explicit. On a Cartesian flat background define
\begin{equation}
 U_v^{\mu\nu}=v^{\mu\rho}\eta_{\rho\lambda}v^{\lambda\nu},
 \qquad \sigma_1^{\mu\nu}=-v^{\mu\nu}+\tfrac12U_v^{\mu\nu}+O(v^3).
 \label{AppDcubicmap}
\end{equation}
Let $S_{3,v}$ and $S_{3,\sigma_1}$ denote the cubic parts of $S+S_{\mathrm{GF}}$ in the two variables, and let $S_2$ be their common quadratic functional. Taylor expansion about the flat vacuum, whose first variation vanishes, gives
\begin{align}
 S_{3,v}[v]-S_{3,\sigma_1}[-v]
 &=-\tfrac12\int\dd^4x\,
       \frac{\delta S_2[v]}{\delta v^{\mu\nu}}U_v^{\mu\nu}\nonumber\\
 &=-\frac1{16\chi}\int\dd^4x\,
       (P(-\Box)v)_{\mu\nu}U_v^{\mu\nu}.
 \label{AppDcubicidentity}
\end{align}
The functional derivative is with respect to the independent contravariant field, and the equality holds for integrated actions with the stated boundary convention. It includes the cubic term generated by $F[\sigma_1(v)]^2$; omitting that term changes the comparison.

For three external free solutions satisfying $P(-\Box)v_i=0$, every term in the multilinear part of the right-hand side vanishes. With the common linear identification $\sigma_{1i}=-v_i$, identical canonical normalization and momentum conservation, the physical transverse-traceless three-point coefficients therefore agree. This is the first interacting-order realization of local field-redefinition equivalence in the chosen gauge~\cite{Kamefuchi1961,Chisholm1961}. The individual off-shell vertices remain different. Equation~\eqref{AppDcubicidentity} is the flat-vacuum identity; on a general off-shell curved background, additional contributions from the first variation must be retained.

Appendix~\ref{AppenB} derives all six density cubic terms, the exponential gauge contribution and Eq.~\eqref{AppDcubicidentity}. Its explicit complex-momentum example evaluates the same nonzero three-point coefficient directly in both variables and matches its normalization to the Einstein-gravity interaction. This checks the interaction without discarding the determinant contribution or introducing a second independent gravitational field.

\section{Schwarzschild configurations and the background-harmonic gauge}
\label{AppE}
We construct a finite Schwarzschild configuration in the background-harmonic gauge of Sec.~\ref{AppB}. The density and exponential variables reconstruct the same metric exactly, and the configuration is stationary under unrestricted variations of the gauge-fixed action in the exterior. We then distinguish this statement from variation within a restricted radial ansatz.

\subsection{Exact density and exponential fields}
Let $m=GM>0$, so that $r_s=2m$. In coordinates $(t,\rho,\theta,\phi)$, choose the fixed flat reference metric
\begin{equation}
 ds_0^2=dt^2-d\rho^2-\rho^2d\Omega^2,
 \qquad d\Omega^2=d\theta^2+\sin^2\theta\,d\phi^2.
 \label{eq:schw-harmonic-background}
\end{equation}
The Schwarzschild exterior has the harmonic-radius representation~\cite{FromholzPoissonWill2014}
\begin{equation}
 ds^2=f(\rho)\,dt^2-f(\rho)^{-1}d\rho^2-(\rho+m)^2d\Omega^2,
 \qquad f(\rho)=\frac{\rho-m}{\rho+m},\qquad \rho>m.
 \label{eq:schw-harmonic-metric}
\end{equation}
Its areal radius is $r=\rho+m$. The associated Cartesian functions
$(t,\rho\sin\theta\cos\phi,\rho\sin\theta\sin\phi,\rho\cos\theta)$ form a harmonic coordinate system. Below we verify the background-covariant gauge directly in spherical coordinates, retaining the nonzero reference connection.

Set $u=m/\rho$, with $0<u<1$. The volume ratio and normalized inverse-density fluctuation are
\begin{align}
 \omega&=\frac{\sqrt{-g}}{\sqrt{-g_0}}=(1+u)^2,\nonumber\\
 \sigma_1^{\mu\nu}
 &=\operatorname{diag}\left(\frac{(1+u)^3}{1-u}-1,\,u^2,\,0,\,0\right).
 \label{eq:schw-exact-fields}
\end{align}
Consequently,
\begin{align}
 Q^\mu{}_{\nu}&=\sigma^{\mu\lambda}g_{0\lambda\nu}
 =\operatorname{diag}\left(\frac{(1+u)^3}{1-u},\,1-u^2,\,1,\,1\right),\nonumber\\
 \det Q&=(1+u)^4=\omega^2,\qquad H=4\ln(1+u).
 \label{eq:schw-harmonic-determinant}
\end{align}
The inverse density is
\begin{equation}
 \tau_{\mu\nu}=\operatorname{diag}\left(
 \frac{1-u}{(1+u)^3},\,-\frac{1}{1-u^2},\,-\rho^2,\,-\rho^2\sin^2\theta\right).
 \label{eq:schw-harmonic-inverse-density}
\end{equation}
Multiplication by $e^{H/2}=\omega$ reproduces every component of Eq.~\eqref{eq:schw-harmonic-metric}.

All eigenvalues of $Q$ are positive on the exterior patch, so its principal real matrix logarithm gives
\begin{equation}
 V^\mu{}_{\nu}=-(\ln Q)^\mu{}_{\nu}
 =\operatorname{diag}\left(\ln\frac{1-u}{(1+u)^3},\,-\ln(1-u^2),\,0,\,0\right).
 \label{eq:schw-harmonic-exponential}
\end{equation}
The nonzero contravariant components are therefore
\begin{equation}
 v^{tt}=\ln\frac{1-u}{(1+u)^3},\qquad
 v^{\rho\rho}=\ln(1-u^2).
 \label{eq:schw-harmonic-v-components}
\end{equation}
In particular, $V^\rho{}_{\rho}=v^{\rho\rho}g_{0\rho\rho}=-v^{\rho\rho}$. Since $\tr V=-H$, the metric exponential is
\begin{equation}
 \mathsf h=V-\tfrac12\Id\,\tr V
 =\operatorname{diag}\bigl(\ln f,\,-\ln f,\,2\ln(1+u),\,2\ln(1+u)\bigr).
 \label{eq:schw-harmonic-h}
\end{equation}
Thus $g=g_0e^{\mathsf h}$ independently reconstructs the same metric. Both fields tend to zero as $\rho\to\infty$, and no weak-field truncation has been used.

\subsection{Exact gauge condition and bulk action}
For the fixed reference metric~\eqref{eq:schw-harmonic-background}, the only potentially nonzero component of $F^\mu=\nab_\nu\sigma_1^{\mu\nu}$ is radial. Since the angular components of $\sigma_1^{\mu\nu}$ vanish, the spherical connection gives
\begin{equation}
 F^\rho=\partial_\rho\sigma_1^{\rho\rho}
 +\frac{2}{\rho}\sigma_1^{\rho\rho}
 =\partial_\rho\left(\frac{m^2}{\rho^2}\right)+\frac{2m^2}{\rho^3}=0,
 \qquad F^t=F^\theta=F^\phi=0.
 \label{eq:schw-harmonic-gauge}
\end{equation}
This is the exact nonlinear condition~\eqref{AppBharmonic}. In the exponential coordinate it is satisfied by $F[\sigma_1(v)]$, with the complete field map retained.

The determinant contribution to the exact density action is nonzero:
\begin{equation}
 H_\rho=-\frac{4m}{\rho(\rho+m)},\qquad
 \frac18\sigma^{\mu\nu}H_\mu H_\nu
 =-\frac{2m^2(\rho-m)}{\rho^4(\rho+m)}.
 \label{eq:schw-harmonic-trace-term}
\end{equation}
Direct evaluation of the other two contractions in Eq.~\eqref{AppAdensitycontractions} gives
\begin{align}
 &-\frac14\sigma^{\mu\nu}\tau_{\alpha\rho}\tau_{\beta\lambda}
 (\nab_\mu\sigma^{\alpha\beta})(\nab_\nu\sigma^{\rho\lambda})\nonumber\\
 &\qquad+\frac12\tau_{\alpha\rho}
 (\nab_\lambda\sigma^{\alpha\beta})(\nab_\beta\sigma^{\lambda\rho})
 =\frac{4m^2}{\rho^3(\rho+m)}.
 \label{eq:schw-harmonic-derivative-terms}
\end{align}
The flatness of $g_0$ then implies
\begin{equation}
 \mathcal L_\sigma=\frac{2m^2}{\rho^4}.
 \label{eq:schw-harmonic-bulk}
\end{equation}
The boundary current in Eq.~\eqref{AppAboundary} is
\begin{equation}
 W^\mu=\left(0,-\frac{2m(\rho-m)}{\rho^3},0,0\right),\qquad
 \nab_\mu W^\mu=\frac1{\rho^2}\partial_\rho(\rho^2W^\rho)
 =-\frac{2m^2}{\rho^4}.
 \label{eq:schw-harmonic-boundary}
\end{equation}
Hence $\mathcal L_\sigma+\nab_\mu W^\mu=0$, consistently with $\omega R[g]=0$. The nonzero bulk density is compatible with the vanishing Ricci scalar because the exact action identity includes this divergence.

\subsection{Unrestricted stationarity}
With primes denoting derivatives with respect to $\rho$, the independent Ricci components of Eq.~\eqref{eq:schw-harmonic-metric} are
\begin{align}
 R_{tt}&=f\left(\frac{f''}{2}+\frac{f'}{\rho+m}\right),&
 R_{\rho\rho}&=-\frac1f\left(\frac{f''}{2}+\frac{f'}{\rho+m}\right),\nonumber\\
 R_{\theta\theta}&=1-f-(\rho+m)f',&
 R_{\phi\phi}&=\sin^2\theta\,R_{\theta\theta}.
 \label{eq:schw-harmonic-ricci}
\end{align}
All off-diagonal components vanish. Substitution of
$f'=2m/(\rho+m)^2$ and $f''=-4m/(\rho+m)^3$ gives $R_{\mu\nu}[g]=0$ throughout the exterior.

At fixed $g_0$, the four-dimensional density reconstruction implies
\begin{align}
 \delta\ln\omega&=\tfrac12\tau_{\alpha\beta}\delta\sigma^{\alpha\beta},\nonumber\\
 \delta g^{\mu\nu}&=\omega^{-1}\left(
 \delta\sigma^{\mu\nu}-\tfrac12\sigma^{\mu\nu}\tau_{\alpha\beta}
 \delta\sigma^{\alpha\beta}\right).
 \label{eq:schw-density-variation}
\end{align}
For arbitrary smooth variations with compact support in the exterior, the Einstein--Hilbert variation therefore becomes
\begin{equation}
 \delta S=-\frac1{4\chi}\int\dd^4x\sqrt{-g_0}\,
 R_{\mu\nu}[g]\,\delta\sigma^{\mu\nu}=0.
 \label{eq:schw-stationarity}
\end{equation}
The gauge-fixing term varies as
\begin{equation}
 \delta S_{\mathrm{GF}}=\frac1{4\chi\alpha}\int\dd^4x\sqrt{-g_0}\,
 F_\mu\nab_\nu\delta\sigma^{\mu\nu},\qquad F_\mu=g_{0\mu\nu}F^\nu.
 \label{eq:schw-gauge-variation}
\end{equation}
Equation~\eqref{eq:schw-harmonic-gauge} consequently establishes $\delta(S+S_{\mathrm{GF}})=0$, in particular at $\alpha=1$. These variations are neither restricted to spherical symmetry nor required to preserve the gauge condition. The locally invertible field map gives the same stationarity in the exponential variable. With vanishing ghost and antighost fields, the configuration is also stationary for the classical ghost-extended action. These statements concern the exterior bulk variational problem; boundary action values require separately specified boundary conditions and terms.

\subsection{Independent radial variation}
For comparison with Appendix~\ref{AppenC}, use the areal-radius ansatz
\begin{equation}
 g_{\mu\nu}=\operatorname{diag}(A,-B,-r^2,-r^2\sin^2\theta),\qquad
 g_{0\mu\nu}=\operatorname{diag}(1,-1,-r^2,-r^2\sin^2\theta),
 \label{eq:schw-metrics}
\end{equation}
where $A(r),B(r)>0$. After angular integration and removal of the radial total derivative displayed in Appendix~\ref{AppenC}, the action is
\begin{equation}
 S_{\mathrm{red}}[A,B]=\frac{2\pi}{\chi}\int dt\,dr\,L_r,
 \qquad L_r=\sqrt{\frac{A}{B^3}}\,[rB'+B(B-1)].
 \label{eq:schw-reduced}
\end{equation}
Primes now denote radial derivatives with respect to $r$. Independent variation gives
\begin{equation}
 rB'+B(B-1)=0,\qquad rA'-A(B-1)=0.
 \label{eq:schw-equations}
\end{equation}
The first equation integrates to $r(1-B^{-1})=2m$, while the pair gives $(\ln AB)'=0$. Asymptotic normalization fixes $A=1-2m/r$ and $B=A^{-1}$. Imposing $A=F(r)$ and $B=F(r)^{-1}$ before variation instead gives
\begin{equation}
 L_r=1-F-rF'=[r(1-F)]',
 \label{eq:schw-premature-restriction}
\end{equation}
whose restricted variation supplies no radial equation. The additional relation $AB=1$ is not a consequence of spherical symmetry alone. Its premature imposition removes an independent variation; this does not constitute a failure of the principle of symmetric criticality~\cite{FelsTorre2002}.

Relative to the flat reference in Eq.~\eqref{eq:schw-metrics}, the areal-radius Schwarzschild representative has
\begin{equation}
 \sigma_1^{\mu\nu}=\operatorname{diag}\left(\frac{2m}{r-2m},\frac{2m}{r},0,0\right),
 \qquad F^\mu=\left(0,\frac{2m}{r^2},0,0\right).
 \label{eq:schw-areal-gauge}
\end{equation}
The harmonic configuration~\eqref{eq:schw-harmonic-metric} uses a different identification of the same physical geometry with the flat reference. A simultaneous coordinate transformation $r=\rho+m$ of both metrics in Eq.~\eqref{eq:schw-metrics} would give background angular coefficients $-(\rho+m)^2$ and $-(\rho+m)^2\sin^2\theta$, rather than those in Eq.~\eqref{eq:schw-harmonic-background}, and would preserve the nonzero gauge vector. The exact gauge-fixed stationary configuration is the representative defined by Eqs.~\eqref{eq:schw-harmonic-background} and \eqref{eq:schw-harmonic-metric}.

\section{Discussion and conclusions}\label{AppF}
We have compared two established choices of metric perturbation, the trace-reversed exponential field $v$ and the inverse-density fluctuation $\sigma_1$, within the same second-order Einstein--Hilbert action. The covariant background split isolates a term linear in $\sigma_1$ and a nonlinear derivative contribution. The exact inverse-density and determinant identities make this a single-variable action in either scheme. The determinant term is part of the gravitational dynamics and is retained throughout.

A single background-divergence gauge, identified exactly with a nonlinear background-harmonic condition, fixes the perturbative construction. Its explicit Faddeev--Popov operator supplies the free ghosts and their interaction with the density field. The curved-background quadratic operators obey Eq.~\eqref{AppIntroHessians}: their Ricci-dependent difference is generated when the nonlinear field map acts on the background tadpole. On Ricci-flat backgrounds they agree under the linear field identification, and their flat propagator contains the usual trace reversal. Including the cosmological term changes the difference to $-A+2\Lambda\Id$, which vanishes at an Einstein saddle. At the first interacting order on a flat background, Eq.~\eqref{AppDcubicidentity} relates the two cubic actions through the free field equation. Its multilinear difference vanishes for the common on-shell external fields, as checked by independent exponential and density calculations in Appendix~\ref{AppenB}, with explicit normalization against the known Einstein-gravity three-point interaction.

The exterior Schwarzschild example uses a finite harmonic representative relative to a fixed flat reference. Both field coordinates reconstruct the metric exactly, the density divergence vanishes, and the complete exterior Ricci tensor is zero. The bulk density and boundary current satisfy the exact action identity. These results establish stationarity of the gauge-fixed action under unrestricted compactly supported exterior variations. The separate areal-radius calculation retains the independent two-function variation: imposing the relation between the radial functions beforehand removes an equation. The distinction between the two reference identifications also explains why a simultaneous coordinate transformation of the original metric pair cannot remove its nonzero gauge vector.

The two schemes organize the same theory differently. Their practical use in a particular higher-order calculation depends on the vertices, gauge and observable involved; no general computational advantage follows from the field redefinition alone. Extending the present classical and tree-level comparison to loops requires a common choice of functional measure, regulator and source prescription, with coordinate Jacobians treated consistently. Those choices are separate from the single Faddeev--Popov gauge construction used here.

\appendix
\section{On the gauge-fixing procedure}\label{AppenA}
The reference metric is fixed in a quantum gauge transformation. Under the convention $\delta_\xi g=\mathcal L_\xi g$, the full metric density transforms as a density, while the reference volume in $\sigma^{\mu\nu}=\sqrt{-g}\,g^{\mu\nu}/\sqrt{-g_0}$ is held fixed. In background-covariant notation,
\begin{equation}
 \delta_\xi\sigma^{\mu\nu}
 =\xi^\rho\nab_\rho\sigma^{\mu\nu}
 -\sigma^{\rho\nu}\nab_\rho\xi^\mu
 -\sigma^{\mu\rho}\nab_\rho\xi^\nu
 +\sigma^{\mu\nu}\nab_\rho\xi^\rho.
 \label{AppenAgauge}
\end{equation}
This is distinct from a simultaneous coordinate transformation of $g$ and $g_0$, under which $\sigma$ is an ordinary tensor. Its linear term at $\sigma_1=0$ is $-\nab^\nu\xi^\mu-\nab^\mu\xi^\nu+g_0^{\mu\nu}\nab\cdot\xi$.

For the single gauge $F^\mu=\nab_\nu\sigma_1^{\mu\nu}$, define
\begin{align}
 N^{\mu\nu}(\sigma_1,c)={}&c^\rho\nab_\rho\sigma_1^{\mu\nu}
 -\sigma_1^{\rho\nu}\nab_\rho c^\mu
 -\sigma_1^{\mu\rho}\nab_\rho c^\nu
 +\sigma_1^{\mu\nu}\nab_\rho c^\rho,\nonumber\\
 (M[\sigma_1]c)^\mu={}&-\Boxz c^\mu-R_0{}^\mu{}_{\rho}c^\rho
                       +\nab_\nu N^{\mu\nu}(\sigma_1,c).
 \label{AppenAoperator}
\end{align}
For example, its field-independent part follows directly from
\begin{align}
 \nab_\nu\delta_c\sigma^{\mu\nu}\big|_{\sigma_1=0}
 &=-\Boxz c^\mu-\nab_\nu\nab^\mu c^\nu
                       +\nab^\mu\nab_\nu c^\nu\nonumber\\
 &=-\Boxz c^\mu-[\nab_\nu,\nab^\mu]c^\nu
 =-\Boxz c^\mu-R_0{}^\mu{}_{\rho}c^\rho.
 \label{AppenAfreeproof}
\end{align}
This explicit operator is the variation of the four-component gauge condition. An additional trace-gradient or determinant condition is neither needed nor included.

Up to a field-independent normalization, the Faddeev--Popov determinant is represented by anticommuting vector fields $c^\mu,\bar c_\mu$ with
\begin{equation}
 S_{\mathrm{gh}}=\int\dd^4x\sqrt{-g_0}\,
                    \bar c_\mu(M[\sigma_1]c)^\mu.
 \label{AppenAghost}
\end{equation}
A constant rescaling of $\bar c$ has been used to choose this normalization. With the Fourier convention of Eq.~\eqref{AppC13}, the flat ghost propagator, omitting the momentum delta function, is
\begin{equation}
 \langle c^\mu(p)\bar c_\nu(-p)\rangle_0
       =\frac{i\delta^\mu{}_{\nu}}{p^2+i0}.
 \label{AppenApropagator}
\end{equation}
Integrating the last term of Eq.~\eqref{AppenAoperator} by parts gives the explicit flat ghost--graviton interaction,
\begin{align}
 S_{\mathrm{gh}}^{(3)}=-\int\dd^4x\,(\partial_\nu\bar c_\mu)
 \bigl[c^\rho\partial_\rho\sigma_1^{\mu\nu}
       -\sigma_1^{\rho\nu}\partial_\rho c^\mu
       -\sigma_1^{\mu\rho}\partial_\rho c^\nu
       +\sigma_1^{\mu\nu}\partial_\rho c^\rho\bigr].
 \label{AppenAcubic}
\end{align}
This specifies the vertex with its derivatives, tensor contractions and sign. On a curved reference metric, replacing these derivatives and the integration measure by their background-covariant versions gives the same interaction. The operator is affine in $\sigma_1$, so the density ghost action has no higher graviton powers. In the exponential scheme one substitutes $\sigma_1(v)$; this generates its higher ghost interactions with the same gauge choice. Functional inverses and determinants must be defined on the chosen domain with any residual zero modes handled separately.

\section{Flat-background verification of the cubic expansion}
\label{AppenB}

This appendix checks the cubic interaction in the two field variables on
a Cartesian flat reference metric.  Write $s^{\mu\nu}\equiv
\sigma_1^{\mu\nu}$, and lower its indices with $\eta_{\mu\nu}$.
The lowered field $s_{\mu\nu}$ is not the perturbation of the inverse
density.  Define
\begin{equation}
 t=\eta_{\mu\nu}s^{\mu\nu},\qquad
 U_{\mu\nu}=s_{\mu\rho}s^\rho{}_{\nu},\qquad
 u=s^{\mu\nu}s_{\mu\nu}.
 \label{eq:B-definitions}
\end{equation}
The expansions needed at cubic order are
\begin{equation}
 \tau_{\mu\nu}=\eta_{\mu\nu}-s_{\mu\nu}+U_{\mu\nu}+O(s^3),
 \qquad
 H=\ln\det(1+s\eta)=t-\frac12u+O(s^3).
 \label{eq:B-inverse-and-trace}
\end{equation}
Here $\tau=\sigma^{-1}$, $\sigma=\eta^{-1}+s$, and
$\chi=4\pi G$ with $\kappa^2=8\chi$.

\subsection{The density cubic interaction}

The cubic terms can be read off without truncating away the determinant contribution. The necessary derivative products are
\begin{align}
 (\partial_\mu\sigma^{\alpha\beta})(\partial_\nu\tau_{\rho\beta})
 &=-\partial_\mu s^{\alpha\beta}\partial_\nu s_{\rho\beta}
   +\partial_\mu s^{\alpha\beta}\partial_\nu U_{\rho\beta}+O(s^4),\nonumber\\
 \sigma^{\mu\nu}\partial_\mu H\partial_\nu H
 &=\partial_\mu t\partial^\mu t
   +s^{\mu\nu}\partial_\mu t\partial_\nu t
   -\partial_\mu t\partial^\mu u+O(s^4).
 \label{eq:B-derivative-products}
\end{align}
In the first line, each quadratic product also multiplies the linear part of the explicit $\sigma$ in Eq.~\eqref{AppA1401}. Collecting all terms of degree three and including the action normalization $-1/(4\chi)$ gives
\begin{align}
 \mathcal L_{3,s}={}&
 \frac1{16\chi}s^{\mu\nu}\partial_\mu s^{\alpha\beta}
                                      \partial_\nu s_{\alpha\beta}
 -\frac1{8\chi}s^{\mu\rho}\partial_\mu s^{\alpha\beta}
                                      \partial_\alpha s_{\rho\beta}
 \nonumber\\
 &-\frac1{16\chi}\partial_\mu s^{\alpha\beta}\partial^\mu U_{\alpha\beta}
 +\frac1{8\chi}\partial^\rho s^{\alpha\beta}\partial_\alpha U_{\rho\beta}
 \nonumber\\
 &-\frac1{32\chi}s^{\mu\nu}\partial_\mu t\partial_\nu t
 +\frac1{32\chi}\partial_\mu t\partial^\mu u.
 \label{eq:B-full-density-cubic}
\end{align}
The last two terms follow from
$\sigma^{\mu\nu}\partial_\mu H\partial_\nu H/8$ in the bulk
Lagrangian and are required off shell.  No transverse-traceless condition
has been used.  The gauge choice of Appendix~\ref{AppenA},
\begin{equation}
 F^\mu[s]=\partial_\nu s^{\mu\nu},\qquad
 S_{\rm GF}[s]=\frac1{8\chi}\int d^4x\,F_\mu F^\mu,
 \label{eq:B-density-gauge}
\end{equation}
is exactly quadratic in $s$ and therefore contributes no cubic term in
this variable.  It gives
\begin{equation}
 S_2[s]=\frac1{16\chi}\int d^4x\,
 s^{\mu\nu}(P(-\Box)s)_{\mu\nu},\qquad
 (Ps)_{\mu\nu}=s_{\mu\nu}-\frac12\eta_{\mu\nu}t.
 \label{eq:B-free-action}
\end{equation}
The trace-reversal operator satisfies $P^2=1$ in four dimensions.

\subsection{Comparison with the exponential variable}

For $V^\mu{}_{\nu}=v^{\mu\rho}\eta_{\rho\nu}$ and
$\sigma=e^{-V}\eta^{-1}$, the field map is
\begin{equation}
 s^{\mu\nu}=-v^{\mu\nu}+\frac12 U_v^{\mu\nu}+O(v^3),
 \qquad U_v^{\mu\nu}=v^{\mu\rho}\eta_{\rho\lambda}v^{\lambda\nu}.
 \label{eq:B-field-map}
\end{equation}
The same gauge is $F[s(v)]$.  Its cubic contribution is explicitly
\begin{equation}
 S_{{\rm GF},v}^{(3)}=-\frac1{8\chi}\int d^4x\,
  (\partial_\nu v^{\mu\nu})(\partial_\lambda U_{v\mu}{}^{\lambda}).
 \label{eq:B-exponential-gauge-cubic}
\end{equation}
It must be retained when comparing the two gauge-fixed cubic actions.
Expanding $S_2[s(v)]+S_{3,s}[s(v)]$ through degree three gives
\begin{align}
 S_{3,v}[v]-S_{3,s}[-v]
 &=-\frac12\int d^4x\,
       \frac{\delta S_2[v]}{\delta v^{\mu\nu}}U_v^{\mu\nu}
 \nonumber\\
 &=-\frac1{16\chi}\int d^4x\,
       (P(-\Box)v)_{\mu\nu}U_v^{\mu\nu}.
 \label{eq:B-cubic-EOM-identity}
\end{align}
This is the derivation of Eq.~\eqref{AppDcubicidentity}, with $S_{3,s}=S_{3,\sigma_1}$. The equality is for integrated gauge-fixed actions with the same boundary convention.
It follows from the quadratic Taylor term in
Eq.~\eqref{eq:B-field-map}; the first variation vanishes on the flat
vacuum.  Thus the difference is proportional to the free field equation.
For three external solutions of that equation, every term in the
multilinear difference vanishes.  With the common linear identification
$s=-v$, the physical transverse-traceless cubic coefficients therefore
agree.  This proves the on-shell cubic comparison without assuming a
pre-existing expression for either three-graviton vertex.

\subsection{Direct exponential expansion}
The exponential interaction can also be calculated directly from the exact density action, without using Eq.~\eqref{eq:B-cubic-EOM-identity}. Set $t_v=\eta_{\mu\nu}v^{\mu\nu}$ and use
\begin{equation}
 \sigma=\eta^{-1}-v+\tfrac12U_v+O(v^3),\qquad
 \tau=\eta+v_{\mathrm{low}}+\tfrac12U_{v,\mathrm{low}}+O(v^3),
 \qquad H=-t_v.
 \label{eq:B-direct-exponential-input}
\end{equation}
In the derivative part of Eq.~\eqref{AppA1401}, the cubic terms containing one derivative of $U_v$ and one derivative of $v$ cancel pairwise upon contracting the background coefficients. For example,
$\partial^\rho U_v^{\alpha\beta}\partial_\alpha v_{\rho\beta}
=\partial^\rho v^{\alpha\beta}\partial_\alpha U_{v\rho\beta}$
by relabeling the contracted indices. The remaining cubic density, including the transported gauge term, is
\begin{align}
 \mathcal L_{3,v}^{\mathrm{tot}}={}&
 -\frac1{16\chi}v^{\mu\nu}\partial_\mu v^{\alpha\beta}\partial_\nu v_{\alpha\beta}
 +\frac1{8\chi}v^{\mu\rho}\partial_\mu v^{\alpha\beta}\partial_\alpha v_{\rho\beta}
 \nonumber\\
 &+\frac1{32\chi}v^{\mu\nu}\partial_\mu t_v\partial_\nu t_v
 -\frac1{8\chi}(\partial_\nu v^{\mu\nu})
                   (\partial_\lambda U_{v\mu}{}^\lambda).
 \label{eq:B-direct-exponential-cubic}
\end{align}
No free equation or tracelessness condition was used here. The last term is precisely Eq.~\eqref{eq:B-exponential-gauge-cubic}; dropping it would change the off-shell comparison.

\subsection{Explicit evaluation of the three-point coefficient}
For an explicit nonzero check, take the following complex momenta and
polarization vectors, in units of a fixed momentum scale:
\begin{align}
 k_1&=(1,0,0,1),& e_1&=(0,1,-i,0),\nonumber\\
 k_2&=(0,1,i,0),& e_2&=(1,0,0,-1),\nonumber\\
 k_3&=(-1,-1,-i,-1),& e_3&=(0,1,i,0).
 \label{eq:B-complex-data}
\end{align}
They satisfy $\sum_i k_i=0$, $k_i^2=e_i^2=k_i\cdot e_i=0$.
Consequently $\epsilon_i^{\mu\nu}=e_i^\mu e_i^\nu$ are transverse
and traceless.  Introduce a common canonically normalized external field
$\phi^{\mu\nu}(x)=\sum_i z_i\epsilon_i^{\mu\nu}e^{ik_i\cdot x}$.
Evaluate the density cubic action at $s=\kappa\phi$ and the exponential
cubic action at $v=-\kappa\phi$. Define $V_3$ as the coefficient of $z_1z_2z_3$ in the action after removing the common momentum-conserving delta function; the vertex is $iV_3$.

Each wave obeys $\Box\phi^{\mu\nu}=0$ and $\partial_\mu\phi^{\mu\nu}=0$. After integration by parts, the third term in Eq.~\eqref{eq:B-full-density-cubic} is proportional to $\Box\phi^{\alpha\beta}$ and the fourth to $\partial_\alpha\partial^\rho\phi^{\alpha\beta}$. Neither contributes. The last two terms vanish because $t=0$. Thus only the first two terms remain. For factorized polarizations their coefficients are
\begin{align}
 V_3^{(1)}&=-\frac{\kappa}{2}
 \sum_{(a,b,c)\in S_3}
 (e_a\cdot k_b)(e_a\cdot k_c)(e_b\cdot e_c)^2,\nonumber\\
 V_3^{(2)}&=\kappa
 \sum_{(a,b,c)\in S_3}
 (e_a\cdot k_b)(e_b\cdot k_c)
 (e_a\cdot e_c)(e_b\cdot e_c).
 \label{eq:B-contraction-formula}
\end{align}
The sum is over the six permutations of the three wave labels, which are not spacetime indices. The signs include the two factors of $i$ from the derivatives. For Eq.~\eqref{eq:B-complex-data}, the relevant contractions are
\begin{equation}
 e_1\cdot e_3=-2,\qquad e_1\cdot e_2=e_2\cdot e_3=0,
 \qquad e_2\cdot k_1=2,\qquad e_2\cdot k_3=-2.
 \label{eq:B-explicit-contractions}
\end{equation}
Only $(a,b,c)=(2,1,3),(2,3,1)$ contribute to $V_3^{(1)}$, and every summand in $V_3^{(2)}$ contains a vanishing polarization contraction. Therefore
\begin{equation}
 V_3^{(1)}=-\frac{\kappa}{2}\,[2(2)(-2)(-2)^2]=16\kappa,
 \qquad V_3^{(2)}=0.
 \label{eq:B-direct-evaluation}
\end{equation}
Independently substitute $v=-\kappa\phi$ into Eq.~\eqref{eq:B-direct-exponential-cubic}. Its trace and gauge terms vanish on these external waves. The first two terms become, respectively,
\begin{equation}
 \frac\kappa2\phi^{\mu\nu}\partial_\mu\phi^{\alpha\beta}\partial_\nu\phi_{\alpha\beta},
 \qquad
 -\kappa\phi^{\mu\rho}\partial_\mu\phi^{\alpha\beta}\partial_\alpha\phi_{\rho\beta}.
 \label{eq:B-exponential-TT-terms}
\end{equation}
Their six-permutation sums are the two expressions in Eq.~\eqref{eq:B-contraction-formula}, giving $16\kappa$ and zero directly. Thus both explicit evaluations give
\begin{equation}
 V_{3,s}=V_{3,v}=16\kappa.
 \label{eq:B-nonzero-check}
\end{equation}
Equivalently, using unit coefficients for the unnormalized $s$ waves
and the opposite $v$ waves gives $16/\kappa^2=2/\chi$.
The displayed value uses the explicit, unnormalized polarization vectors
in Eq.~\eqref{eq:B-complex-data}; restoring a momentum scale $E$
multiplies it by $E^2$.  Complex momenta allow a nonzero three-point
coefficient, which real collinear massless momenta would not test.

\subsection{Normalization against the Einstein-gravity three-point interaction}
For an external benchmark, use the cubic Einstein-gravity Lagrangian in Eq.~(10) of Bern and Grant~\cite{BernGrant1999}. Denote its canonically normalized graviton by $q_{\mu\nu}$ to distinguish it from the exponential metric variable $h$ used here. Its linear metric normalization is $g_{\mu\nu}=\eta_{\mu\nu}+\kappa q_{\mu\nu}+O(q^2)$, with the same $\kappa^2=32\pi G$ and signature $(+---)$. Their Weinberg curvature convention has the opposite sign to Eq.~\eqref{AppA2}, so their $+2\sqrt{-g}R/\kappa^2$ is our Einstein--Hilbert action. With indices restored, their cubic expression is
\begin{equation}
 \mathcal L_{3,\mathrm{GR}}[q]
 =\kappa\left[\tfrac12q^{\mu\nu}q^{\rho\sigma}
       \partial_\mu\partial_\nu q_{\rho\sigma}
       +q^{\nu\mu}(\partial^\lambda q^\rho{}_{\mu})
                       (\partial_\nu q_{\rho\lambda})\right].
 \label{eq:B-GR-reference-cubic}
\end{equation}
This is an on-shell benchmark in the reference's field and gauge conventions, not an identification of its full off-shell action with ours. We evaluate the displayed Lagrangian with our Fourier waves $e^{ik\cdot x}$ and the vertex convention $iV_3$; consequently each pair of derivatives supplies $i^2=-1$.

For $q=\phi$ and arbitrary factorized transverse null polarizations, its two terms give
\begin{align}
 V_{3,\mathrm{GR}}[\phi]={}&
 -\frac\kappa2\sum_{(a,b,c)\in S_3}(e_a\cdot k_b)^2(e_b\cdot e_c)^2
 \nonumber\\
 &-\kappa\sum_{(a,b,c)\in S_3}
 (e_b\cdot k_c)(e_c\cdot k_a)(e_a\cdot e_b)(e_a\cdot e_c)
 =-\kappa J^2,\nonumber\\
 J={}&(e_1\cdot e_2)(e_3\cdot k_1)
      +(e_2\cdot e_3)(e_1\cdot k_2)
      +(e_3\cdot e_1)(e_2\cdot k_3).
 \label{eq:B-GR-contractions}
\end{align}
The last equality uses momentum conservation and $e_i\cdot k_i=0$. The corresponding sums in Eq.~\eqref{eq:B-contraction-formula} give $+\kappa J^2$. The apparent sign difference is fixed by the external-field identification: $s=\kappa\phi$ and $v=-\kappa\phi$ both give $g=\eta-\kappa\phi_{\mathrm{low}}+O(\phi^2)$ for traceless waves. Therefore the matching reference field is $q=-\phi_{\mathrm{low}}$, and
\begin{equation}
 iV_{3,s}=iV_{3,v}=iV_{3,\mathrm{GR}}[-\phi]
 =i\kappa J^2=16i\kappa
 \quad\hbox{for Eq.~\eqref{eq:B-complex-data}}.
 \label{eq:B-GR-match}
\end{equation}
No polarization rescaling or coupling is stripped in this equality: each external tensor is exactly $e_i^\mu e_i^\nu$ with the field sign just specified, and the common $(2\pi)^4\delta^{(4)}(\sum_i k_i)$ is omitted. Rescaling $e_i$ by $c_i$ multiplies every result by $\prod_i c_i^2$. As a further check, replacing any $e_i$ by $k_i$ makes $J$ vanish using $k_i\cdot k_j=0$. Multilinearity then implies invariance under $e_i\mapsto e_i+a_i k_i$, providing the on-shell Ward check. These are validations of the known Einstein-gravity interaction.

The analytic comparison includes the full off-shell cubic actions, their field-redefinition identity, direct on-shell evaluations in both variables and the normalization check above. As additional algebraic checks, the metric reconstructed from $\sigma$ was differentiated directly to form its Christoffel symbols and Ricci scalar, and its quadratic and cubic coefficients were compared with the density expressions using exact symbolic arithmetic. The direct exponential formula was also checked for traceful off-shell fields, and the amplitude comparison for gauge-shifted and complex-conjugate external data. 

\section{Radial curvature, variation and boundary identities}
\label{AppenC}
This appendix supplies the algebra underlying the two-function variation in Sec.~\ref{AppE}. For the metric in Eq.~\eqref{eq:schw-metrics}, the nonzero independent Ricci components computed with Eq.~\eqref{AppA2} are
\begin{align}
 R_{tt}&=\frac{A''}{2B}
 -\frac{A'}{4B}\left(\frac{A'}{A}+\frac{B'}{B}\right)
 +\frac{A'}{rB},\nonumber\\
 R_{rr}&=-\frac{A''}{2A}
 +\frac{A'}{4A}\left(\frac{A'}{A}+\frac{B'}{B}\right)
 +\frac{B'}{rB},\nonumber\\
 R_{\theta\theta}&=1-\frac1B
 +\frac{r}{2B}\left(\frac{B'}B-\frac{A'}A\right),
 \qquad R_{\phi\phi}=\sin^2\theta\,R_{\theta\theta}.
 \label{eq:schw-two-function-ricci}
\end{align}
Their contraction gives
\begin{equation}
 r^2\sqrt{AB}\,R
 =\left(\frac{r^2A'}{\sqrt{AB}}\right)'-2L_r,
 \label{eq:schw-radial-boundary}
\end{equation}
where $L_r$ is defined in Eq.~\eqref{eq:schw-reduced}. After angular integration, the prefactor of the left-hand side in the Einstein--Hilbert action is $-\pi/\chi$. Removing the displayed radial total derivative therefore yields $S_{\mathrm{red}}=(2\pi/\chi)\int dt\,dr\,L_r$.

The independent Euler--Lagrange derivatives are
\begin{equation}
 \mathcal E_A(L_r)=\frac{L_r}{2A},\qquad
 \mathcal E_B(L_r)=\frac{\sqrt{AB}}{2AB^2}
       [A(B-1)-rA'],
 \qquad
 \mathcal E_A=\frac{\partial}{\partial A}
       -\frac{\dd}{\dd r}\frac{\partial}{\partial A'},
 \label{eq:schw-independent-variations}
\end{equation}
with the analogous definition of $\mathcal E_B$. Setting them to zero for $A,B>0$ gives Eq.~\eqref{eq:schw-equations}.

The restriction $A=F$, $B=F^{-1}$ instead gives $L_r=1-F-rF'=[r(1-F)]'$. It also sets $\omega=1$ and $H=0$. Evaluating the connection difference in Eq.~\eqref{AppAconnection}, the boundary current in Eq.~\eqref{AppAboundary} has only the radial component
\begin{equation}
 W^r=F'+\frac{2(F-1)}r,
 \qquad R=\frac1{r^2}(r^2W^r)',
 \qquad \mathcal L_\sigma=0.
 \label{eq:schw-restricted-boundary}
\end{equation}
The second identity follows directly by contracting Eq.~\eqref{eq:schw-two-function-ricci} after the restriction. The last then follows from $\omega R=\mathcal L_\sigma+\nab_\rho W^\rho$ and the reference volume element. These identities hold for arbitrary $F(r)$ on the nonsingular patch, not only for Schwarzschild. They explain why the restricted action cannot determine the Schwarzschild profile; the independent variations in Eq.~\eqref{eq:schw-independent-variations} are required.

\bibliography{perturbation}
\end{document}